# Scintillation decay-time constants for alpha particles and electrons in liquid xenon

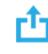

View Online    Export Citation    CrossMark

Dominick Cichon,[a] 🔟 Guillaume Eurin,[b] 🔟 Florian Jörg, 🔟 Teresa Marrodán Undagoitia, 🔟 and Natascha Rupp

**AFFILIATIONS**

Max-Planck-Institut für Kernphysik, Saupfercheckweg 1, 69117 Heidelberg, Germany

[a] Author to whom correspondence should be addressed: dominick.cichon@mpi-hd.mpg.de
[b] Current address: IRFU, CEA, Université Paris-Saclay, F-91191 Gif-sur-Yvette, France.

**ABSTRACT**

Understanding liquid xenon scintillation and ionization processes is of great interest to improve analysis methods in current and future detectors. In this paper, we investigate the dynamics of the scintillation process for excitation by $\mathcal{O}(10 \text{ keV})$ electrons from a $^{83m}$Kr source and $\mathcal{O}(6 \text{ MeV})$ $\alpha$-particles from a $^{222}$Rn source, both mixed with the xenon target. The single photon sampling method is used to record photon arrival times in order to obtain the corresponding time distributions for different applied electric fields between about 0.8 V cm$^{-1}$ to 1.2 kV cm$^{-1}$. Energy and field dependencies of the signals, which are observed in the results, are discussed.

## I. INTRODUCTION

Over the last years, liquid xenon detectors have demonstrated great potential for the direct detection of dark matter particles[1–4] and for the search for the neutrinoless double-beta decay of $^{136}$Xe.[5] For single phase detectors, position reconstruction of an interaction and background rejection both depend solely on properties of the prompt scintillation pulse, called S1, which results from an energy deposition. For instruments operated as dual phase time projection chambers (TPCs), which are also able to measure a charge signal induced by ionization electrons, called S2, the S1 pulse plays a central role in the modeling of the detector response. A better understanding of photon generation in liquid xenon (LXe) is, therefore, crucial to model the response of LXe to ionizing radiation. Improving this modeling can help to achieve better rare-event search sensitivities with upcoming LXe detectors.[6–9]

One specific avenue is to utilize the scintillation pulse shape. Techniques based on scintillation pulse shape information have already been employed in a variety of experiments. In noble gas detectors searching for particle dark matter, such as liquid argon detectors and the aforementioned LXe detectors, the pulse shape enables a separation between nuclear recoils (NRs), which, for example, could be induced by weakly interacting massive particles (WIMPs) scattering off nucleons, and electronic recoils (ERs) originating from background radiation.[10–12] While detectors that utilize both charge and prompt scintillation signals usually employ the ratio between the two for ER/NR separation, pulse shape information can be combined with that ratio to achieve an even better separation.[13,14] In the case of LXe TPCs operating at drift field strengths smaller than 100 V cm$^{-1}$, the improvement is significant.[15] This technique is also used with organic scintillators to separate alpha from beta particles[16,17] or to identify special signatures.[18]

A particle interacting with LXe deposits energy in three different ways: ionization, excitation, and elastic collisions. The elastic collisions produce no measurable quanta, but heat the medium instead. Excited xenon atoms combine with ground state xenon atoms to form excited dimers (excimers), which decay under the emission of vacuum ultraviolet (VUV) radiation, with the emission spectrum being centered around 175 nm.[19] The time scale of the decay depends on whether the electron is in a spin singlet or a spin triplet state (several ns vs several tens of ns). Consequently, two different decay-time constants have been typically observed in the past. A fraction of the ionization electrons recombines with parent ions, resulting in additional excimers which contribute to the prompt scintillation signal. Electron recombination depends on the strength of the applied electric field. The stronger the field, the easier it is to extract electrons from the interaction site, preventing their recombination.

The recombination process has a typical time constant as well. Its effect on the liquid xenon scintillation pulse shape depends not only on the applied field, but also on the amount of energy deposited per unit length by a particle ($dE/dx$). The larger $dE/dx$, the denser the distribution of ions and excited xenon atoms within the particle's track, which affects the recombination and energy loss processes. It is this dependence which allows us to distinguish between species of interacting particles, for instance $\alpha$-particles and electrons or NRs and ERs, based on the pulse shape alone. $dE/dx$ also affects the measured effective singlet and triplet state lifetimes. For example, the singlet and triplet lifetimes for $\mathcal{O}(<1$ MeV$)$ ER scintillation signals have been measured to lie around 2.5 and 25 ns, respectively, if they happen under an electric field of $\mathcal{O}(100$ V cm$^{-1})$.[14,20–22] In the case of $\alpha$-particles, only data at zero field are available, which give roughly 4 and 24 ns, respectively.[23–25]

In this work, we aim at determining the time evolution of the prompt scintillation process in LXe for both $\mathcal{O}(10$ keV$)$ ERs induced by electrons from the isomeric transition (IT) of $^{83m}$Kr[26] and $\mathcal{O}(6$ MeV$)$ $\alpha$-particles from the decay of $^{222}$Rn and its daughters[27] at different electric fields. Photon arrival times distributions have been measured at more than 20 different field strengths between ~0.8 V cm$^{-1}$ and ~1.2 kV cm$^{-1}$, with the focus being on the low-field range (<200 V cm$^{-1}$). Most field configurations have been measured twice to investigate potential systematic uncertainties. To extract pulse shape parameters, an effective model for the photon emission probability is then fit to the measured photon arrival time distributions. Our measurement of $\alpha$-particles is the first which has been conducted in such detail. The extracted model parameters are afterwards interpreted in the context of previously published data.

## II. EXPERIMENTAL METHOD

For measuring photon arrival times, the method of single photon sampling[28] is used, which is outlined in the following and illustrated in Fig. 1. LXe scintillation pulses are observed by two photomultiplier tubes (PMTs). The first one is set up to register as many photons per pulse as possible. The second one is behind an attenuator with the goal of observing, on average, less than one photon per pulse seen in the first PMT. As the signal detected by the first PMT contains a large number of photons, the time at which the scintillation process starts can be determined with high accuracy. Using the time difference between the scintillation start time as measured by the first PMT and the corresponding single photons observed in the second PMT, called $\Delta t$ in the following, the probability density function (PDF) for the photon arrival time can be determined. As this technique does not rely much on accurate detector response modeling, it is expected to be robust against corresponding systematic errors.

The setup used for the measurements presented in this publication is the Heidelberg Xenon (HeXe) xenon TPC, operated in single-phase mode, which has previously been used for other published studies.[29–31] The LXe target is enclosed by polytetrafluoroethylene (PTFE) in a cylindrical volume of 5.6 cm diameter and 5 cm height. PTFE is chosen due to its excellent reflective properties at the peak scintillation wavelength of LXe.[32,33] Two PMTs are placed on top and bottom of the volume (see Fig. 1) to detect scintillation photons originating from it. The PMTs are R6041-406 Hamamatsu tubes of 2 in. diameter with a specified transit time spread of only 0.75 ns (FWHM).

A PTFE attenuator is placed in front of the top PMT to reduce the average amount of photons seen, as required by the single photon sampling method. In addition, an attenuator is also placed in front of the bottom PMT if the amount of scintillation light detected would result in saturation of either the PMT itself or the signal electronics, which is the case for the measured $\alpha$-decays. Required attenuator thicknesses for the signals from each of the radionuclides measured have been estimated using preliminary results from a separate measurement,[34] together with in situ measurements of the amount of light seen by each PMT. For $^{83m}$Kr, 2 mm of PTFE with a ~200 $\mu$m diameter central pinhole has been placed in front of the top PMT. For the $^{222}$Rn measurement, PTFE attenuators have been put in front of both the top and bottom PMT. The top attenuator is

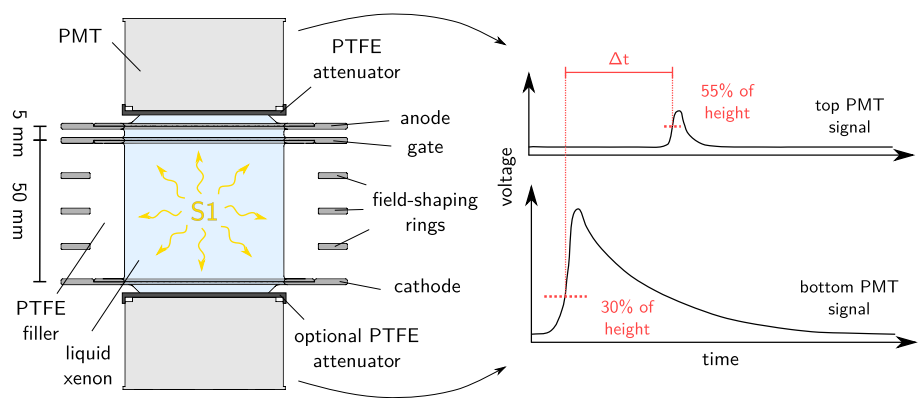

**FIG. 1.** Illustration of the HeXe TPC in single-phase mode and the single photon sampling method. Recoiling electrons and $\alpha$-particles interact with the LXe, which results in prompt scintillation signals (S1s) that are observed by two PMTs. An attenuator made out of PTFE is located in front of the top PMT to reduce the average amount of photons seen per S1 to $\mathcal{O}(1)$ or less as required by single photon sampling. The bottom PMT signal is used to determine a reference point that is fixed relative to the start of the scintillation process, while the top PMT signal allows us to measure the arrival time of single photons relative to that reference point. Reasons for the choice of using the top and bottom PMT signal reference points as shown here for defining $\Delta t$ are given in Sec. IV.

~2.5 mm thick and has a central pinhole of ~300 $\mu$m, which is about 2 mm deep. The bottom attenuator is ~700 $\mu$m thick.

PMT signals are voltage-amplified via a custom-built NIM module (factor ~10 for top PMT and factor ~2 for bottom PMT signals). They are also split by the module and fed into both a digitizer and a discriminator. The former is a CAEN V1743[33] switched capacitor digitizer, which samples at a frequency of 3.2 GHz and has a dynamic range of 2.5 V at 12 bit resolution. The latter is a Phillips Scientific 711 discriminator, which sends a trigger to the digitizer as soon as the bottom PMT signal passes a certain threshold. The threshold is chosen such that its impact on the acceptance for the signals to be measured is negligible.

The photosensors and signal processing electronics are all expected to contribute to the $\Delta t$ uncertainty. Converting the PMT transit time spread to a standard deviation gives an absolute lower limit of 0.32 ns for the achievable time resolution. This would be the case if the bottom PMT were to see an infinite amount of photons, and no other sources of statistical uncertainty were present. As the number of photons seen by the bottom PMT is finite, an additional uncertainty contribution is expected due to both photon statistics and the bottom PMT transit time spread. An indirect contribution is added by both the amplifier and digitizer, as their parameters affect how precise signal reference points, for example a point in the signal's rising edge, can be determined. Furthermore, the choice of the reference point itself is crucial when it comes to its statistical variance and robustness against noise. For this reason, a procedure for selecting the best possible reference points for both top and bottom PMT signals is outlined in Sec. IV.

The electric field inside the LXe target is created by applying specified voltages to the TPC's anode, gate, and cathode. The electrodes are hexagonal meshes with a pitch of 1 mm and a thickness of 120 $\mu$m and have been chosen to achieve good field homogeneity within the drift volume,[31] which is the region between the gate and cathode. The field strength has been varied during a measurement over more than 20 different values between ~0.8 V cm$^{-1}$ and ~1.2 kV cm$^{-1}$ to determine its effect on the scintillation pulse shape. Each field configuration has been measured for at least 1 h, with most configurations having been measured twice at different points in time to assess the reproducibility of our results.

## III. ELECTRIC FIELD ESTIMATION

In order to estimate both strength and direction of the field generated by the electrode grids, the setup's geometry is re-created within a three-dimensional COMSOL Multiphysics® simulation.[31] The simulation output is used to ensure optimum field homogeneity within the volume between the gate and cathode, which contain most of the LXe target. Because each PMT is operated at a fixed voltage, the presence of the PMTs does not allow achieving the same field in all detector regions simultaneously. The resulting inhomogeneities between detector subvolumes (for example, below the cathode or above the anode) introduce a systematic uncertainty to the target field strength. This uncertainty depends on the relative amount of photons detected in the top PMT from each subvolume. The simulated electric field as a function of the location within the TPC, $(r^2, z)$, is shown in part (a) of Fig. 2 for a drift volume field of about 1 kV cm$^{-1}$.

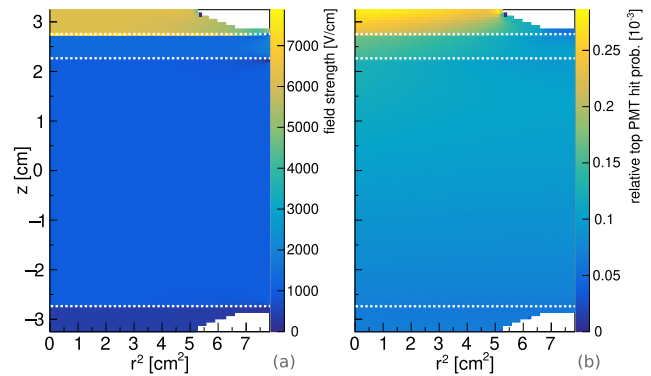

**FIG. 2.** (a) Simulated electric field within the TPC for a target field strength of ~1 kV cm$^{-1}$. The map has been averaged over the azimuthal angle with respect to the central axis. The central 90% of the drift volume field strength distribution (central volume between the gate and cathode) is within 2.5% of the region's median. (b) Simulated top PMT hit probability map, normalized to yield an integral of 1. Both plots shown are for the attenuator configuration used for measuring $^{222}$Rn data. The white, dashed lines mark the electrode grid locations.

The probability of scintillation photons to be detected in the top PMT depends on the location of the event within the TPC. This is due to the light collection efficiency (LCE) being affected by the detector geometry and the optical properties of the detector materials, such as both reflectivity and transmittance of PTFE. In order to estimate this position-dependent detection probability, an optical Geant4[36] Monte Carlo simulation, which includes the entire TPC geometry, together with the PTFE attenuators, is carried out. The simulation produces photons isotropically and homogeneously distributed within the LXe target and keeps track of which photons hit the PMT photocathodes. For the optical parameters of each material, the same values as in Ref. 34 are used. The final result is an LCE map [shown in part (b) of Fig. 2] which contains the position-dependent probability of detecting a photon in the top PMT. The pinhole of the top PMT attenuator used for the $^{222}$Rn measurement does not visibly affect the LCE map, as most photons seen by the top PMT pass directly through the attenuator material.

To get a precise estimate for the field strength of each field configuration, the TPC volume is divided into several voxels with identical volumes. For each voxel, the average field strength within it is determined using the output of the electric field simulation. To take the photon detection probability into account, the voxels are weighted with the average LCE for interactions happening within them, as extracted from the optical Monte Carlo. All voxels together then yield a field strength distribution. This distribution, under the assumption of interactions being homogeneously distributed within the TPC, translates to the probability of an interaction having occurred while being subjected to a certain field strength, given that a single photon has generated a signal in the top PMT. For the central value of a field strength estimate, the distribution's median is used, while the interquartile range (IQR) is chosen as a measure for its systematic uncertainty. The latter is, in this case, equivalent to a field strength interval within which 50% of the observed photons in the top PMT have been emitted under. For the field configurations of the measurements presented here, the IQR typically ranges within $\mathcal{O}(1 \text{ V cm}^{-1})$. As is evident from Fig. 2, detector subvolumes other than the drift volume still need to be accounted for when interpreting the results, as the IQR is not sensitive to the tails of the distribution.

## IV. ANALYSIS AND METHODS

The recorded data are first corrected by the HeXe data processor[37] at the raw waveform level for artifacts caused by the V1743 digitizer and for crosstalk induced by the bottom PMT channel into the top PMT one. Digitizer artifacts result from the time-interleaving method used to achieve its sampling frequency of 3.2 GHz by recording an input signal on 16 different lines in parallel. These lines sample at 200 MHz each, and are delayed by 312.5 ps relative to each other. If the gains and offsets of these lines do not match, the result is an artificial pattern with a period of 16 digitizer samples, which depends on the input signal size. While the factory calibration of the digitizer should eliminate this issue, it is found to be not entirely sufficient in our case. The pattern is corrected for mainly by measuring it at different input voltages. Then, a model is built, which models the pattern via cubic spline interpolation, after which the resulting estimated pattern can be subtracted from the corresponding raw waveform. The model accounts for both gain and offset mismatches.

Crosstalk is modeled by fitting a digital filter to the top PMT waveforms, which only contain crosstalk while using the corresponding bottom PMT waveforms as input. Fit samples are obtained by switching off the top PMT while measuring scintillation signals from $^{222}$Rn (daughter) $\alpha$-decays with the bottom one. The fitted model then allows us to estimate and subtract crosstalk from the top PMT waveform. After all corrections are applied, the data processor identifies scintillation pulses and calculates parameters such as area and height. Detailed descriptions of the raw data corrections mentioned above and the signal identification algorithms are given in Ref. 37.

Event samples for the analysis are defined by applying several selection criteria. For all measurements, we require that the event happens during stable field conditions and that only a single signal is found in the top PMT channel. That signal needs to have an area between 0.1 times and 0.75 times the average single photoelectron response for $^{83m}$Kr data and between 0.1 times and 1.0 times the average single photoelectron response for $^{222}$Rn data. This requirement effectively introduces a selection on the detector volume as the geometrical photon detection probability is not uniform. Its effects are accounted for by the electric field estimation as outlined in Sec. III.

In the case of $^{83m}$Kr, the delayed coincidence between the initial 32.2 keV IT of the nuclide followed by a 9.4 keV IT with a half-life of ~154 ns[18] can be utilized.[31,39] This is done by selecting events where two signals are found in the bottom PMT, with the start of the second-largest signal as defined by the data processor having to occur at least 160 ns after the start of the largest signal. It also ensures that within an interval of about 150 ns, only photons from the 32.2 keV decay contribute to the $\Delta t$ distribution. In addition, the area of the second-largest signal is histogrammed against the area of the largest one to fit a two-dimensional Gaussian via a least squares minimization (see Fig. 3). This is done separately for each field configuration because of the ER light yield dependence on field strength not being negligible.[31,39,40] An additional criterion is then applied, where events need to be in the central 85% of the Gaussian, while the smaller of the two signals is larger than a lower threshold, which is given in Fig. 3. This is to remove events where the two decay signals cannot be separated from each other and are coincidentally followed by a bottom PMT dark pulse.

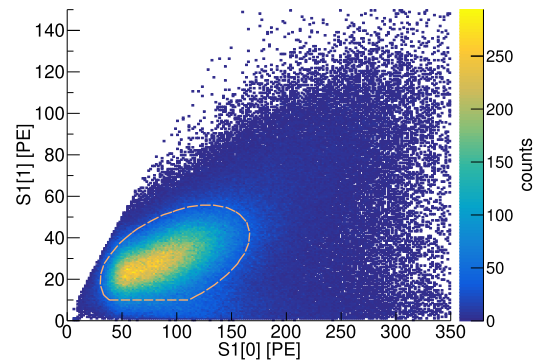

**FIG. 3.** Area of second-largest signal (S1[1]) vs area of largest one (S1[0]) in units of average PMT signal area per photoelectron (PE) at a field of ~1 keV cm$^{-1}$. The dashed line indicates the selection region, which is defined by the central 85% probability region of a fitted two-dimensional Gaussian and a lower threshold on S1[1], which is 10 PE.

For $^{222}$Rn $\alpha$-decays and the ones of its daughters $^{218}$Po and $^{214}$Po are selected by requiring that only a single signal is found in the bottom PMT. Also, the signal's area needs to be within a certain region of the scintillation signal spectrum. Because the light yield dependence of the investigated $\alpha$-decays is negligible within the measured field strength range,[31,41] the selection region is the same for each field configuration. It has been verified that this region actually contains the aforementioned $\alpha$-decays using dual-phase data from a series of preliminary measurements for the analysis presented in Ref. 31. The latter contain both scintillation and ionization signals, which together allow to separate the decays from each other. This is not possible when only using the scintillation signal due to insufficient energy resolution in that case. Besides the requirement on the area, a final requirement is made on the bottom PMT signal height-to-area ratio in order to remove events from background radiation and cosmic muons, which are identified by comparing the selection space before and after the injection of $^{222}$Rn.

To define $\Delta t$ for the analysis, reference points on the rising or the falling edge of a signal that are independent of the signal's size need to be selected for both top and bottom PMT signals. In the top PMT, where single photoelectron signals are employed, the point of steepest gradient is chosen as a reference. It is determined by least squares fitting an exponentially modified Gaussian[42] to $\mathcal{O}(1000)$ top PMT waveforms, which contain such a signal. According to the extracted model parameters, the point of steepest gradient is approximately equivalent to the time in the rising edge of a signal where 55% of the signal's height is reached. For this reason, the latter is determined by the data processor as the top PMT signal reference point. For bottom PMT signals, which consist of multiple photoelectrons, the time in the rising edge where 30% of the signal height is reached is used to define the reference point.

The resulting $\Delta t$ distribution is modeled by convolving an effective model probability density function for the scintillation emission time with a Gaussian distribution, which results in a sum of two exponentially modified Gaussian distributions

$$p(\Delta t) = f_s p_s(\Delta t) + (1 - f_s)p_t(\Delta t), \quad (1)$$

with:

$$p_{s/t}(\Delta t) = \frac{1}{2\tau_{s/t}} e^{\frac{1}{2}\left(\frac{\sigma_t}{\tau_{s/t}}\right)^2} e^{\frac{-(\Delta t - t_0)}{\tau_{s/t}}} \operatorname{erfc}\left[\frac{1}{\sqrt{2}\sigma_t}\left(\frac{\sigma_t^2}{\tau_{s/t}} - (\Delta t - t_0)\right)\right]. \quad (2)$$

This effective model, which has been used by previous publications in either this form or a similar one,[11,12,14,21,22] consists of two exponential decays with decay constants $\tau_s$ and $\tau_t$ that correspond to the singlet and the triplet state of a xenon excimer, respectively. $f_s$ denotes the fraction of scintillation photons emitted from singlet state decays compared to the total number of singlet and triplet decay photons. The Gaussian distribution is used to model the $\Delta t$ uncertainty via its standard deviation $\sigma_t$, which is referred to as the time resolution in the following. The choice of a Gaussian is supported by the $\Delta t$ distributions having a left flank, which is shaped correspondingly, as seen in Fig. 4. $t_0$ is an offset parameter to account for timing differences between top and bottom PMT signals. These can be caused by the signal electronics and by biases resulting from the particular choice of the PMT signal reference points.

The effective model is fit to the measured $\Delta t$ distributions via unbinned negative log-likelihood minimization. In the case of $^{83m}$Kr data, a first iteration of fits with Eq. (1) is conducted for each field configuration in order to determine a value and statistical error for $\sigma_t$. Afterward, $\sigma_t$ is fixed to its average over all field configurations during a second, final iteration of fits with the same function in order to improve fit stability and mitigate correlation effects between $\tau_s, f_s$, and $\sigma_t$. The actual model parameter results are taken from that final iteration. For $^{222}$Rn data, it is found that a third exponential decay needs to be added to the model in order to describe the measured distributions:

$$\tilde{p}(\Delta t) = (1 - f_3)p(\Delta t) + f_3 p_3(\Delta t), \quad (3)$$

with $p(\Delta t)$ corresponding to Eq. (1) and $p_3(\Delta t)$ corresponding to Eq. (2), but with $\tau_3$, which denotes the decay constant of the third component instead of $\tau_{s/t}$. $f_3$ is the total fraction of photons contributed by the third component. The origin of this additional component is currently unknown, with possible causes being discussed in Sec. V. To enhance fit stability, which is important when repeating the fit many times for estimating systematic uncertainties later on, $\tau_3$ is constrained via another method. This is done by designating a late region in the $\Delta t$ distribution, in which the third component contributes at least ~90% of all photons (Fig. 4). This fraction, and thus the size of this region are estimated using parameter values extracted from preliminary fits. A simple exponential decay is then fit to the region in order to extract both value and statistical error for $\tau_3$. Then, Eq. (3) is fit to the rest of the distribution to obtain a value and statistical error for $\sigma_t$. After that, $\tau_3$ and $\sigma_t$ are averaged over all field configurations and kept fixed at these values. Finally, a second iteration analogous to the $^{83m}$Kr case is made to extract the model parameters while excluding the late region.

Systematic uncertainties introduced by the $\Delta t$ fit region bounds, the upper limit on the top PMT signal area, and the particular choice of the central 2D Gaussian region in case of $^{83m}$Kr data are estimated by varying them within ±5%–30% of their value depending on the parameter. The range of variation is chosen such that it is assumed to capture systematic deviations, while not being so large that the uncertainty estimate is artificially increased. For example, the upper limit on the top PMT signal area is varied for $^{222}$Rn data by a large range of about 30%, as no prior assumptions regarding a safe upper bound can be made except for the fact that it should be smaller than the average single photoelectron signal size. Statistical uncertainties of $\sigma_t$ and $\tau_3$ are accounted for by varying them within their 68.3% confidence level interval. Systematic uncertainties for $\tau_3$ are estimated by repeating the late region fit while varying the other parameters mentioned above. In the end, the systematic error is taken for each fit parameter as the central 68.3% interval of the distribution of all fit results obtained during variation. We note that the systematic uncertainties for each data point produced by this method are correlated with each other.

For $^{222}$Rn data, the lower $\Delta t$ fit region bound is not varied in order to ensure that all fits converge during systematic uncertainty estimation. This is necessary due to residual digitizer artifacts, which remain even after raw waveform correction, and negatively affect fit stability.

Only data points for which every single fit during systematic error estimation succeeded are accepted. This yields results for 22 different field configurations out of 23 with $^{83m}$Kr data. For $^{222}$Rn data, 22 field configurations out of 30 remain for $\tau_3$, while 26 remain for the other model parameters.

## V. RESULTS AND DISCUSSION

Figure 4 shows the measured $\Delta t$ distributions for ERs induced by the 32.2 keV decay of $^{83m}$Kr at the highest and lowest field strengths measured, together with an example $^{222}$Rn measurement. The fits of the effective model to each distribution are also shown, together with indicators for the fit region and the late region used to determine $\tau_3$ in the $^{222}$Rn measurement. The left edge of the distribution is partially excluded from the fit. This choice has been made because of digitizer artifacts, which remain even after raw waveform correction. During preliminary fits, an oscillatory pattern is observed in the fit residuals, with minima and maxima being apart from each other by multiples of roughly 5 ns, respectively, 16 digitizer samples (see Fig. 5). This implies that the artificial time-interleaving pattern is not entirely eliminated. The residual pattern, in turn, biases the determination of the signal reference points. In order to mitigate biases in the resulting fit parameter, the left bound of each measurement's fit interval is selected to approximately lie at a zero-crossing of the oscillatory pattern. This method improves the fit stability and reduces the bias for parameters that are constrained by the left edge of the $\Delta t$ distribution, especially $\tau_s$, $f_s$, and $\sigma_t$, which are highly correlated with each other ($r \sim 0.8$ and larger).

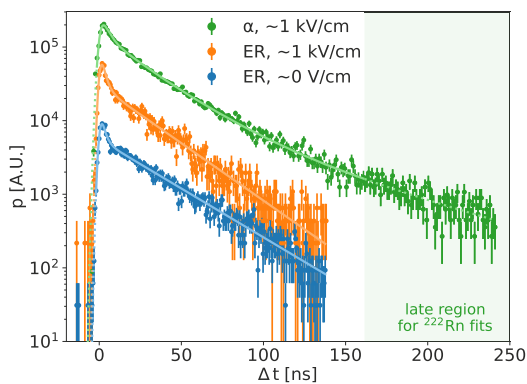

**FIG. 4.** Data of three $\Delta t$ probability density functions, taken at ~0.8 V cm$^{-1}$ and ~1 kV cm$^{-1}$ electric field strength, together with the final fits of the effective model to them. They have been shifted on the x-axis by $-t_0$ to align them based on the scintillation start time and on the y-axis for better visibility. The fit function is drawn with a solid line within the interval that is used for fitting. Shading indicates the late region used to determine $\tau_3$ in $^{222}$Rn data fits.

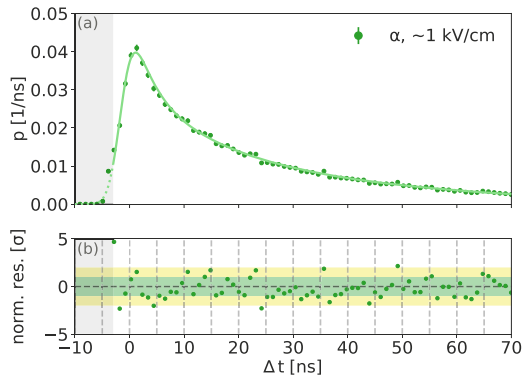

**FIG. 5.** (a) Zoomed, linear y-axis view of the $\alpha$-particle $\Delta t$ distribution from Fig. 4 without any axis shifts. The final fit is also shown. (b) Normalized residuals of the final fit. The vertical dashed lines are 5 ns apart from each other and illustrate that the distances between maxima and minima of the residuals are close to multiples of 5 ns, which is equivalent to 16 digitizer samples.

The results for $\sigma_t$ obtained before the second iteration of fits are shown in Fig. 6. For both $^{83m}$Kr and $^{222}$Rn data the resolution remains, in good approximation, constant over the entire field range. This supports the choice of fixing it for the final iteration of fits. The fact that the measured resolution is worst for $^{222}$Rn data is attributed to the residual digitizer pattern affecting larger signals more than smaller ones.

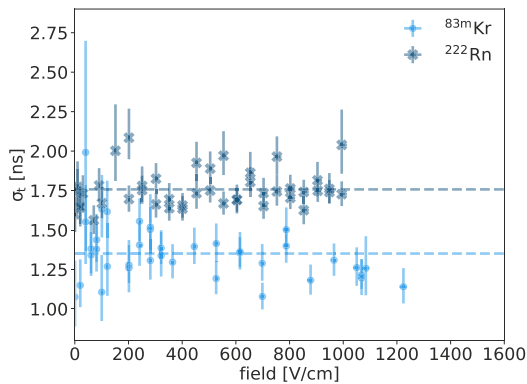

**FIG. 6.** Results for $\sigma_t$ after the first iteration of fits for both $^{83m}$Kr and $^{222}$Rn data. Average values, which are indicated by dashed lines, are $(1.35 \pm 0.03)$ and $(1.757 \pm 0.017)$ ns, respectively.

Results for ERs from the 32.2 keV $^{83m}$Kr decay are shown in Fig. 7. Data points of field configurations that have been measured twice are compatible with each other. They are also generally in agreement with the published values that are shown in the figure. These values have been derived using the same or an equivalent effective pulse shape model at similar energies and fields, but via

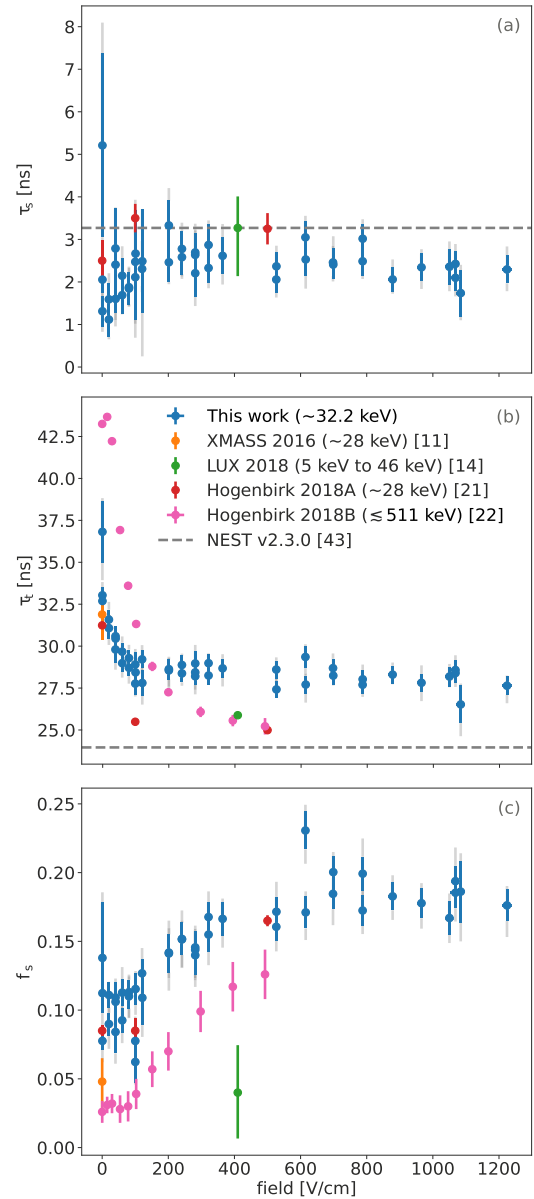

**FIG. 7.** Pulse shape parameter results for ERs induced by the 32.2 keV $^{83m}$Kr IT. Error bars for electric field and statistical parameter errors are solid, while error bars for systematic parameter errors are transparent and gray. Existing literature data points for ERs at either similar energies or a similar field range are shown for comparison. They correspond to XMASS 2016,[11] LUX 2018,[14] Hogenbirk 2018A,[21] and Hogenbirk 2018B.[22] The singlet and triplet state lifetimes as assumed by NEST v2.3.0[43] are also shown.

Monte Carlo matching instead of single photon sampling. Also shown for comparison are the values used in the Noble Element Simulation Technique (NEST) software package v2.3.0[43] for simulating prompt scintillation signals of ERs in LXe. The singlet and triplet lifetimes used by NEST correspond to values measured in Ref. 14, with the triplet lifetime being taken from a result for NRs even though an ER result is also available in the same publication. When comparing results, one has to keep in mind that NEST models photons from excimers produced by direct excitation in a different way compared to photons from excimers produced by recombination when simulating ER signals (see also Ref. 44). In the latter case, the time needed until recombination is modeled as well. Furthermore, NEST assumes different singlet fractions for direct excitation excimers and recombination excimers. For this reason, no direct comparison is made for that parameter.

Results for $\tau_s$ and $f_s$ match with the ones at similar ER energies, with the only disagreement being with the significantly lower $f_s$ value published by the LUX collaboration.[14] For $f_s$, the observed dependence on the applied field qualitatively agrees with the Hogenbirk *et al.* measurement for ERs induced by 511 keV gamma photons scattering off LXe.[22] Quantitatively, the $f_s$ values measured in that publication are systematically smaller than the ones from this work. This can be explained by the difference in the field between the measurements. Previous publications indicate that $f_s$ depends on the linear energy transfer (LET) of the particle interacting with the LXe, with $f_s$ being larger for a larger LET.[21,25] It is hypothesized in Ref. 25 that this is a result of singlet state excimers undergoing superelastic collisions with thermal electrons, in which they are scattered into triplet excimers. As the density of the ionization track is larger for higher LET values, recombination occurs correspondingly faster with less time available for ionization electrons to partake in such collisions and change singlet into triplet state excimers. The ESTAR database[45] cites a xenon stopping power of about 4.7 MeV cm$^2$ g$^{-1}$ for 32.2 keV electrons, while it is roughly a factor 3.5 lower for ~340 keV Compton edge electrons produced by 511 keV gamma scattering.

For $\tau_s$, a downward trend toward lower field strengths is visible, which starts at ~250 V cm$^{-1}$ and is compatible with Ref. 21. It coincides with the measured rise of $\tau_t$ toward lower field strengths, which qualitatively matches the already known behavior of the effective triplet state lifetime, which has been observed to decrease with increasing electric field.[20,22,46] Both trends are attributed to recombination dynamics, which affect the photon emission time for excimers produced by recombination and become more noticeable toward weaker fields. The results at the lowest measured field strength are compatible with the other measurements at similar ER energies. The disagreement with Ref. 22 around that field range could be explained by the difference in ER energy and, thus, LET. As mentioned above, the $\mathcal{O}(100 \text{ keV})$ electrons used for that publication for measuring the ER scintillation pulse shape have a significantly lower LET than the ones at $\mathcal{O}(10 \text{ keV})$ used in this work. This implies a lower ionization track density for the former, which would imply recombination to take longer compared to the latter.

The limit value of $\tau_t$ toward larger field strengths (about 28 ns) is significantly larger than the ~25 ns implied by the other publications, which are all compatible with each other when disregarding NEST, which uses a different pulse shape model. A possible explanation is field inhomogeneities within the HeXe TPC. When optimizing for field homogeneity between the gate and cathode, the voltages applied to the cathode tend to be similar to the one the bottom PMT is operated at (see Fig. 2 for an example). This results in a region with a field close to zero, with events occurring within it being able to affect the $\Delta t$ distribution.

The order of magnitude of the resulting bias is estimated using the weighted field distributions from Sec. III for sampling the electric field and modeling the field dependence of $f_s$ and $\tau_t$ using the results from this work and the ones from Ref. 22 and fitting polynomials to them. $f_s$ is modeled using the results from this work. For $\tau_t$, results from Ref. 22 are taken above 190 V cm$^{-1}$, which is the point at which the field evolution measured there and the one measured in this work cross over. Below that field, results from this work are used. The field samples are then translated into values for $\tau_t$ and $f_s$, which are then used together with the other $^{83m}$Kr results from this work to sample a photon detection time using Eq. (1). In total, $10^5$ photon detection times are sampled. Afterward, an unbinned negative log-likelihood fit is done with Eq. (1) to compare the fit parameter results with the input values, and the $f_s$ and $\tau_t$ values at the median field strength of the field configuration, respectively. The entire procedure is repeated 100 times per field configuration, and yields a $\tau_t$ bias of up to +5% toward high fields and −5% for field strengths lower than 200 V cm$^{-1}$. This indicates that roughly half of the difference between both measurements at high fields could be accounted for by field inhomogeneities.

Another point to consider when comparing values from this work to those from Ref. 22 are differences in fit methodology. The latter fixes both $\tau_s$ and $\sigma_t$ to values that have been fitted in a previous publication,[21] while we let $\tau_s$ vary freely. This causes variations in the pulse shape to be absorbed by different parameters, meaning that a field dependence of $\tau_s$ as seen in this analysis could be absorbed by $\tau_t$ and $f_s$ in the analysis from Ref. 22.

In the case of $\mathcal{O}(6 \text{ MeV})$ $\alpha$-particles from $^{222}$Rn and its daughters, the $\tau_3$ values obtained in the first iteration of fits (see Fig. 8) are constant over the entire field range, with data points measured twice being compatible with each other. This also holds true for the other effective pulse shape parameters that are shown in Fig. 9. This is consistent with the interpretation of any field dependence being caused by recombination dynamics. The relative change in light yield observed within the field range investigated here amounts to at most ∼1%,[31,41] which leads to the expectation of a correspondingly small influence of recombination changes on the scintillation pulse shape. This matches observations from Ref. 46, which measured a constant area-over-amplitude ratio for $\alpha$-particles from the decay of $^{241}$Am for fields from 0 to 3.8 kV cm$^{-1}$.

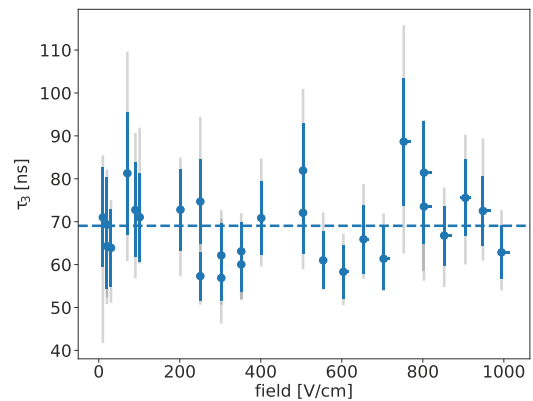

**FIG. 8.** Results for $\tau_3$ after the first iteration of fits for $^{222}$Rn data. The average value is indicated by a dashed line and corresponds to $(69.0 \pm 1.8_{\text{stat}}{}^{+5}_{-5\,\text{sys}})$ ns.

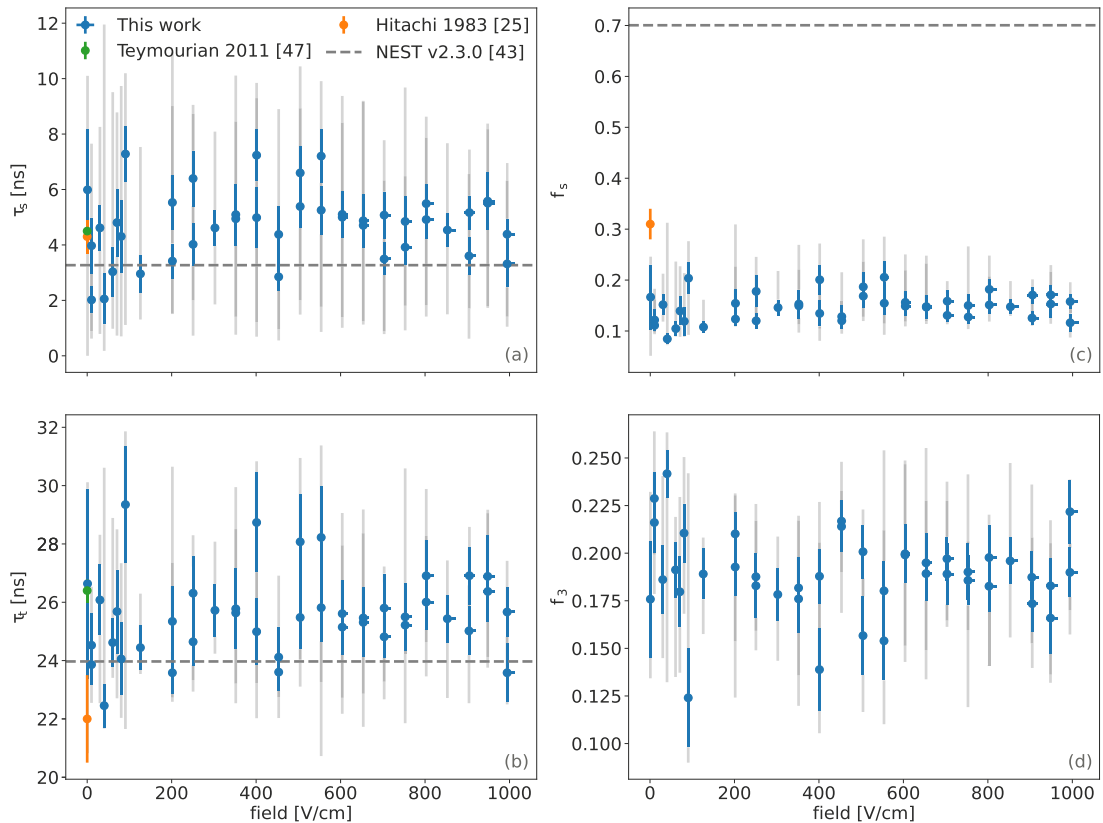

**FIG. 9.** Pulse shape parameter results for $\mathcal{O}(6\text{ MeV})$ $\alpha$-particles from the decays of $^{222}$Rn, $^{218}$Po, and $^{214}$Po. Error bars for electric field and statistical parameter errors are solid, while error bars for systematic parameter errors are transparent and gray. Existing literature data points for $\alpha$-particles are shown for comparison. They correspond to Hitachi 1983[25] ($^{210}$Po and $^{252}$Cf) and Teymourian 2011[47] ($^{210}$Po). The singlet/triplet state lifetime and singlet fraction assumed by NEST v2.3.0[43] for 5.5 MeV $\alpha$-particles are also shown.

The sparse amount of literature on the $\alpha$-particle pulse shape employs $\Delta t$ distribution models with two components equivalent to Eq. (1), while the analysis in this work needs to add a third component to fit the measured photon arrival time distributions [Eq. (3)]. Still, the values extracted for $\tau_s$ and $\tau_t$ are in agreement with previously published values (Fig. 9 and Refs. 23 and 24). In the case of $\tau_s$, this is already true when only accounting for statistical errors. This hints toward $\tau_s$, indeed, being roughly twice as large compared to scintillation pulses induced by ERs and also hints toward differences in the microphysics processes responsible for populating the excimer states. The $\tau_s$ and $\tau_t$ values used by NEST for $\alpha$-particles are the same as those used for ERs and are compatible with the measurements in this work and the presented literature data.

The measured singlet fraction $f_s$ is comparable to the one measured for $^{83m}$Kr at fields above 500 V cm$^{-1}$. This is incompatible, however, with the superelastic collision hypothesis, according to which one would expect a larger singlet fraction for $\alpha$-particles than for $\mathcal{O}(10\text{ keV})$ ERs as the former have, according to the ASTAR database,[45] a stopping power which is two orders of magnitude larger compared to the latter. The value of $0.31 \pm 0.03$ from Ref. 25, in contrast, supports the hypothesis. One also has to consider systematic effects, which affect the comparison of the effective pulse shape parameters. For example, the fit in Ref. 25 includes a constant offset, which can be seen as also accounting for a third component, as observed in our work. Approximately modeling a third exponential decay via a constant would affect the extracted $f_s$ value in a different way compared to using a different model. For instance, the constant could absorb more of the triplet component compared to the singlet one, as the distribution of the former is more similar to a uniform distribution compared to the latter. This would result in a larger singlet fraction value being measured. However, because the constant would need to absorb roughly half of the triplet contribution for such an effect to completely account for the discrepancy, it is unlikely that the only cause is the difference in model functions. A significantly larger deviation is visible when comparing with the $f_s$ value used by NEST for 5.5 MeV $\alpha$-particles (approximate $\alpha$-particle energy of $^{222}$Rn decay[27]), which is off by more than a factor of 2 compared to the other results. The NEST value closely matches and is likely related to the one given in Ref 44, which has been determined using a fit to data from Refs. 46 and 47. No details regarding the fit procedure are available. Thus, no statement about potential causes for the NEST deviation can be made here.

The origin of the third component, which makes up about 19% of all scintillation photons, is currently unknown. Photon delays due to travel within the PTFE attenuators seem unlikely as no third component is observed in the $^{83m}$Kr data. Photosensor light emission or similar effects are considered to be unlikely as well, as the amount of potential photons emitted from a PMT would need to be at the same order of magnitude as the amount of $\alpha$-particle interaction scintillation light in order to be seen at a similar intensity in the PMT opposite to it. Furthermore, as the $\Delta t$ distribution is measured using events with only one single photon signal in the top PMT, effects that create an additional delayed signal, such as afterpulsing, do not affect our measurement. Also, such effects should appear at a similar size in both $^{83m}$Kr and $^{222}$Rn data. The remaining causes of the third component are either speculative xenon microphysics processes, which become relevant at high LET stopping power only ($> 100$ MeV cm$^2$ g$^{-1}$), or impurities within the HeXe TPC that take part in the microphysics. For the latter, LET-dependent energy transfer to N$_2$ might be a candidate as the HeXe TPC is kept in a pure nitrogen atmosphere every time it is opened. While the cryostat housing the HeXe TPC is evacuated for at least a day before filling and the xenon is continuously purified during operation, the PTFE filler material, which is in direct contact with the xenon, could store and release N$_2$ over time so slowly, that it cannot be efficiently cleaned. Because the parameters of the third component were stable over the course of the measurement, this would also necessitate N$_2$ to have been emanated over the entire measurement duration, which amounts to roughly a week.

A final summary of the pulse shape parameter results for both particle sources is given in Table I. $^{222}$Rn values are averaged over the entire measured field range, while $^{83m}$Kr values are averaged above 500 V cm$^{-1}$ as they remain constant in that range within the sensitivity of this analysis.

**TABLE I.** Effective liquid xenon pulse shape parameter values for $^{83m}$Kr (ERs induced by 32.2 keV IT) and $^{222}$Rn ($\alpha$-particles with roughly 6 MeV) data, averaged over field configurations above 500 V cm$^{-1}$ for the former and over all field configurations for the latter. The $\tau_t$ values for $^{83m}$Kr shown here include the approximate systematic error due to field inhomogeneities, which is estimated to be as large as $-5\%$, but is not as precisely determined as other systematic effects.

| | Data | |
|---|---|---|
| Parameter | $^{83m}$Kr (E $\geq$ 500 V cm$^{-1}$) | $^{222}$Rn |
| $\tau_s$ (ns) | $2.38 \pm 0.09$ stat $^{+0.16}_{-0.12}$ sys | $4.73 \pm 0.14$ stat $^{+3}_{-3}$ sys |
| $\tau_t$ (ns) | $28.09 \pm 0.16$ stat $^{+0.3}_{-1.7}$ sys | $25.56 \pm 0.19$ stat $^{+2}_{-2}$ sys |
| $\tau_3$ (ns) | $\cdots$ | $69.0 \pm 1.8$ stat $^{+5}_{-5}$ sys |
| $f_s$ (%) | $18.4 \pm 0.3$ stat $^{+0.6}_{-0.9}$ sys | $14.7 \pm 0.4$ stat $^{+6}_{-2}$ sys |
| $f_3$ (%) | $\cdots$ | $18.9 \pm 0.2$ stat $^{+3}_{-3}$ sys |

## VI. CONCLUSIONS

We have studied pulse shape parameters for prompt scintillation light emitted by LXe as a function of electric field for the first time consistently for both excitation by $\mathcal{O}$(10 keV) ERs and $\alpha$-particles of MeV energies. To measure scintillation photon arrival time distributions, the single photon sampling method has been employed. ER data from the 32.2 keV decay of $^{83m}$Kr and $\mathcal{O}$(6 MeV) $\alpha$-particle data from the decays of $^{222}$Rn, $^{218}$Po, and $^{214}$Po are employed. For that purpose, the aforementioned radionuclides have been introduced into the HeXe liquid xenon TPC operated in single-phase mode. After the measurements, an effective model consisting of either two or three exponential decays, which correspond to the singlet, respectively, the triplet xenon excimer state decays and an additional component, has been fit to the photon arrival time distributions. Systematic uncertainties that affect the electric field and the fit results have been studied in detail. The electric field under which the measured photons have been emitted has been estimated using a three-dimensional simulation while, simultaneously, accounting for the photon detection probability, which has been extracted from an optical Monte Carlo simulation of the detector. Systematic uncertainties resulting from the event selection and fitting method have been estimated by simultaneously varying selection parameters and fit region bounds.

The results for $^{83m}$Kr ERs show a dependence on the field strength, which matches previously published data. Also, they agree with the superelastic collision hypothesis, which predicts a larger amount of singlet state decays compared to triplet state ones for larger particle LETs.

Results for $^{222}$Rn, $^{218}$Po, and $^{214}$Po $\alpha$-particles, which represent, to our best knowledge, the first comprehensive measurement of effective pulse shape parameters for LXe scintillation induced by $\alpha$-particles over a wide electric field range, indicate a stable pulse shape between ~0.8 and ~1000 V cm$^{-1}$. Previously published data exist only for zero field, and our values of $\tau_s$ and $\tau_t$ are compatible with those. There is a discrepancy when it comes to the $\alpha$-particle value for $f_s$ from Ref. 25 and the ER measurements when assuming the superelastic collision hypothesis to be true, which predicts $f_s$ to be larger for the former compared to the latter. A potential explanation for this is differences in the effective model used to extract the pulse shape parameters. The origin of the third component requires further investigation, with impurities within the LXe target being a potential candidate. Microphysics processes which are unaccounted for could also play a role.

## ACKNOWLEDGMENTS

We acknowledge the support of the Max Planck Society (Crossref funder registry entry: https://doi.org/10.13039/501100004189). We thank our technicians Hannes Bonet, Steffen Form, Michael Reißfelder, and Jonas Westermann for their support during construction and operation of the Heidelberg Xenon system. We also thank Michael Reißfelder, in particular, for the manufacturing of the PTFE attenuators, which have been essential for the measurements presented in this work.

## AUTHOR DECLARATIONS

### Conflict of Interest

The authors have no conflicts to disclose.

### Author Contributions

**Dominick Cichon**: Conceptualization (equal); Data curation (equal); Formal analysis (lead); Writing – original draft (lead); Writing – review & editing (equal). **Guillaume Eurin**: Conceptualization (supporting); Data curation (equal); Writing – original draft (supporting); Writing – review & editing (supporting). **Florian Jörg**: Conceptualization (supporting); Data curation (equal); Writing – original draft (supporting); Writing – review & editing (supporting). **Teresa Marrodán Undagoitia**: Conceptualization (equal); Formal analysis (supporting); Supervision (lead); Writing – original draft (supporting); Writing – review & editing (supporting). **Natascha Rupp**: Conceptualization (supporting); Data curation (equal); Writing – original draft (supporting); Writing – review & editing (supporting).

## DATA AVAILABILITY

The data that support the findings of this study are available from the corresponding author upon reasonable request.